\documentclass{aa}
\usepackage[varg]{txfonts}
\usepackage{graphicx}
\usepackage{natbib}
\usepackage{xcolor}
\usepackage{float}
\usepackage{pdflscape}
\usepackage{threeparttable}
\usepackage{multicol}
\usepackage{subcaption}
\usepackage{pgf}
\usepackage[linkcolor=black,citecolor=cyan,filecolor=black,urlcolor=magenta]{hyperref}
\title{The [CII] 158 $\mu$m emission line as a gas mass tracer in high redshift quiescent galaxies} 

\author{C.~D'Eugenio \inst{1,2} 
 \and E.~Daddi\inst{3} 
 \and D.~Liu\inst{4}
 \and R.~Gobat\inst{5}
 }

\institute{Instituto de Astrofísica de Canarias, C. Vía Láctea, s/n, 38205 La Laguna, Tenerife, Spain
\and
Universidad de La Laguna, Dpto. Astrofísica, 38206 La Laguna, Tenerife, Spain
\and 
Université Paris-Saclay, Université Paris Cité, CEA, CNRS, AIM, 91191, Gif-sur-Yvette, France
\and 
Max Planck Institute for Astronomy, Königstuhl 17, D-69117 Heidelberg, Germany
\and
Instituto de F\'isica, Pontificia Universidad Cat\'olica de Valpara\'iso, Casilla 4059, Valpara\'iso, Chile\\
}

\offprints{C. D'Eugenio, \email{cdeugenio@iac.es}, \email{cdeugenio1@gmail.com}}

\begin{document}

\date{Received 19/06/2023; Accepted 15/09/2023}

\abstract{Many efforts have been done in recent years to probe the gas fraction evolution of massive quiescent galaxies (QGs); however, a clear picture has not yet been established. Recent spectroscopic confirmations at z>3 offer the chance to measure the residual gas reservoirs of massive galaxies a few hundreds of Myr after their death and to study how fast quenching proceeds in a highly star-forming Universe. Even so, stringent constraints at z$>$2 remain hardly accessible with the Atacama Large Millimeter/submillimeter Array (ALMA) when adopting molecular gas tracers commonly used for the quenched population. 
In this letter, we propose overcoming this impasse by using the carbon [CII] 158 $\mu$m emission line to systematically probe the gaseous budget of unlensed QGs at z>2.8, when these galaxies could still host non-negligible star formation on an absolute scale and when the line becomes best observable with ALMA (Bands 8 and 7).
So far predominantly used for star-forming galaxies, this emission line is the best choice to probe the gas budget of spectroscopically confirmed QGs at $z>3$, reaching 2-4 and 13-30 times deeper than dust continuum (ALMA band 7) and CO(2-1)/(1-0) (Very Large Array (VLA) K-K$\alpha$ bands), respectively, at fixed integration time. Exploiting archival ALMA observations, we place conservative 3$\sigma$ upper limits on the molecular gas fraction (f$_{\rm{mol}}=M_{H_2}/M_{\star}$) of ADF22-QG1 (f$_{\rm{mol}}$<21\%), ZF-COS-20115 (f$_{\rm{mol}}$<3.2\%), two of the best-studied high-z QGs in the literature, and GS-9209 (f$_{\rm{mol}}$<72\%), the most distant massive QG discovered to date. The deep upper limit found for ZF-COS-20115 is 3 times lower than previously anticipated for high-z QGs suggesting, at best, the existence of a large scatter in the f$_{\rm{mol}}$ distribution of the first QGs. Lastly, we discuss the current limitations of the method and propose ways to mitigate some of them by exploiting ALMA bands 9 and 10.
}

\maketitle

\section{Introduction}
Disentangling the different processes that lead to the formation of massive quenched galaxies is one of the most complex tasks in galaxy evolution \citep{ManBelli18}.
A common aspect to all quenching mechanisms is that they directly affect the amount of cold gas available for star formation. A deep characterisation of the interstellar medium in QGs is therefore necessary. If near-infrared spectroscopic campaigns allowed the stellar component of QGs to be studied up to z$\sim$4.7 with impressive data quality \citep{Carnall23}, our view of their gaseous reservoir becomes increasingly less clear already beyond the local Universe due to the general faintness of this galaxy population to commonly adopted H$_{2}$ proxies, such as dust continuum or low-J CO transitions \citep{Sargent15, Bezanson19, Williams21}. To date, no general consensus has been reached on the residual cold gas fraction of massive QGs at 1<z<3 which is broadly constrained from a few per cent to around 10-20\%, primarily due to rather limited statistics, in addition to uncertainties in different conversion factors, sample selection criteria and intrinsic scatter \citep{Williams21, Whitaker21b, blanquez23}. 
At higher redshifts, our chances to study the gaseous component of individual unlensed QGs drop due to the loss of depth of the aforementioned gas tracers. 
Perhaps the most attainable task is constraining the relevance of bulge growth, which is expected to decrease the star formation efficiency of galaxies whose cold gas fraction dropped below $\sim$20\% keeping them not entirely devoid of cold gas \citep{Martig09, Gobat18}. This, however, requires tracers bright enough to break such a threshold as a function of redshift at reasonable observational costs.
Critically, ALMA’s sensitivity in band 6 would make it unpractical to probe the dust mass or low-J CO transitions at depths comparable to those reached in \citealt{Whitaker21b} in high-z unlensed galaxies until the advent of the next-generation VLA. Notably, low-J CO transitions in QGs will also be difficult to detect against the increasing Cosmic Microwave Background (CMB): 
assuming local thermal equilibrium and following the results of \citealt{Magdis21} at z=2, the kinetic temperature of CO can be expected to follow the decrease in size of QGs with redshift \citep{lustig}, which would lead to T$_{\rm{kin}}\sim$24K at z$\sim$3. In this case, from Eq.(32) in \citealt{dacunha13}, the recoverable fraction of CO(1-0) flux would be $\sim$60\% the intrinsic one at z=3. The loss of contrast is even worse in the case of no intrinsic gas temperature evolution with redshift.
To date, therefore, understanding whether fast-quenching channels at z$>$3 efficiently deplete more than 80-90\% of the gas of massive galaxies remains limited to the availability of lensed galaxies or far-IR stacks \citep{Whitaker21b, Gobat18, Magdis21, suzuki}. This prevents us from fully accessing an epoch in which massive QGs start to appear and when the effects of fast quenching processes can be best studied due to the less likely relevance of slow quenching channels at early cosmic times. New and complementary methods are therefore needed.

In this letter we propose probing the residual gas content of QGs at z$>$3 by extending to this galaxy population the use of the [CII] emission line at 157.74 $\mu$m as a cold gas tracer, to date primarily adopted for star-forming galaxies at z$>$4. We discuss the advantages of the method in Sect.~\ref{sec:Pros}. We describe the three massive QGs at z>2.8 with publicly available coordinates and archival [CII] ALMA coverage in Sect.~\ref{sec:selection}. We then apply the method in Sect.~\ref{sec:obs}, yielding upper limits on their cold
gas fraction. Finally, in Sect.~\ref{sec:discussion} we interpret the results obtained and discuss the potential caveats related to using this tracer for the quenched population.
We assume a $\Lambda$CDM cosmology with H$_{0}=70$ km s$^{-1}$ Mpc$^{-1}$, $\Omega_{M}=0.27$, $\Omega_{\Lambda}=0.73,$ and a \citet{Chabrier03} initial mass function (IMF). Magnitudes are given in the AB photometric system.

\section{Studying high-z quiescent galaxies through [CII]}
\label{sec:Pros}
In order to secure robust spectroscopic confirmations with pre-JWST telescopes, QGs at z$\geq$3 have been generally selected with a magnitude cut in either H or K band (22 and 22.5 AB mag, respectively), thus primarily selecting recently quenched or transitioning systems with modest dust reddening \citep{Schreiber18a, Valentino20, Deuge20, Forrest20a, Forrest2020b, Saracco20}.  Their specific star formation rate (sSFR) is typically required to be less than 10\% that of the coeval Main Sequence (MS), and therefore, as the normalisation of the MS increases towards high-z, the sSFR of many high-z QGs can be comparable to that of local MS galaxies \citep[e.g.][]{Wuyts11b}, while still being quenched compared to the typical coeval star-forming galaxy. This alone suggests that they likely host [CII] emission. Notably, redder UVJ or NUVrJ QGs could host significantly more dust at fixed stellar age, increasing the pool of sources detectable with [CII].\\ Several studies have argued that the bulk of $^{2}P_{3/2}$--$^{2}P_{1/2}$ [CII] fine structure emission originates from photo-dissociation regions (PDRs) in the external layers of molecular clouds heated by the far-UV radiation emitted from hot stars \citep[e.g.][]{Stacey91, Cormier15, Wolfire22}. \citealt{Zanella18} (hereafter Z18) first proposed adopting this emission line as a molecular gas tracer, having shown that the [CII] luminosity (L$_{\rm{[C II]}}$) tightly correlates with the molecular gas mass of star-forming galaxies from z=0 up to z$\sim$6 regardless of their elevation on the MS. This correlation holds over 4 dex in luminosity and with an L$_{\rm{[C II]}}$-to-H$_{2}$ conversion factor ($\alpha_{\rm{[CII]}} \sim31 \,M_{\odot}/L_{\odot}$) which varies by a factor of two at most when galaxies of different star formation efficiencies (SFE=SFR/M$_{\rm{mol}}$) are considered \citep[Z18,][]{Sommovigo21}. 

\citealt[]{Lapham2017} (hereafter L17) show a power-law correlation between the CO(1-0) and [CII] fluxes in a sample of 20 nearby gas-rich early-type galaxies (ETGs) from the Atlas3D sample \citep{Young11}. This correlation shares a similar slope and scatter as for late-type galaxies (LTGs), showing a trend that holds over almost 3 dex in line luminosities. This demonstrates the existence of a correlation in the local Universe between [CII] and CO(1-0) luminosities in galaxies with sSFRs as low as sSFR$< 3\cdot 10^{-11} yr^{-1}$ and SFRs less than a few M$_{\odot}\,yr^{-1}$. This supports the hypothesis for which the Z18 relation would also hold for high-z QGs because the correlation in L17 arises from galaxies with SFRs lower than  can be expected from the bulk of massive QGs z$>$2.8. Moreover, this correlation ensures the presence of [CII] emission also in galaxies with severely suppressed star formation at high-z. Remarkably, this also implies that LTGs and ETGs share, on average, the same $\alpha_{\rm{[CII]}} /\alpha_{\rm{CO}} $ ratio, supporting the use of [CII] as a molecular gas mass tracer also in the quenched population.

If the exact contribution of different gaseous components to L$_{\rm{[C II]}}$ is still discussed, at least 45\% of its luminosity is arising from the molecular component \citep[e.g. Z18;][]{Tarantino21}, allowing to place conservative upper limits. In this regard, L17 also notes that the average fraction of L$_{\rm{[C II]}}$ arising from PDRs and not from purely ionised regions is 63.5-70.9\% for their sample ETGs, hence the large majority. Ionisation fractions larger than half were found in 25\% of their sample, possibly associated with AGN activity. This average fraction is similar to that of spiral galaxies in their work and to other fractions previously reported in the literature for star-forming galaxies \citep[e.g.][]{Stacey91, Cormier15}.

The reliability of the Z18 calibration between L$_{\rm{[C II]}}$ and M$_{H_2}$ at z$>$4 has recently been further tested by \citealt{DZ20} by using 11 ALPINE galaxies at 4$<$z$<$6. By converting L$_{\rm{[C II]}}$ into L$^{\prime}_{\rm{CO(1-0)}}$ using the Z18 relation and an $\alpha_{\rm{CO}} $=4.36 M$_{\odot}$(K km s$^{-1}$ pc$^{2}$)$^{-1}$, all the galaxies considered were found to lie within the 0.38 dex dispersion of the expected relation between L$^{\prime}_{\rm{CO(1-0)}}$ and L$_{\rm{IR}}$ for MS and starburst galaxies in place between z$\sim$0 and z$\sim$5.3. An average systematic overestimation by a factor of 2 was found when comparing L$_{\rm{[C II]}}$ to single-band dust continuum measurements. This offset was reported to be dependent on the scaling used for dust continuum and its origin remains difficult to attribute to either one of the two tracers. The average $\alpha_{\rm{[CII]}} $ from Z18 and \citealt{DZ20} are consistent with those of massive galaxies (M$_{\star}>10^9 M_{\odot}$) at z$\sim$6 from the SIMBA simulations, where the [CII] emission is almost entirely arising from the molecular phase \citep{Vizgan22}. 
Although the relative contributions from HI and H$_{2}$ to L$_{\rm{[C II]}}$ are still debated \citep[e.g.][]{Heintz21} and the calibration between [CII] and other molecular gas tracers still suffers from limited statistics at z$>$2, we note that CO and dust continuum as H$_{2}$ proxies are also affected by increasingly larger uncertainties with increasing redshift (i.e. surface brightness dimming, decrease in gas-phase metallicity, uncertainty on the dust temperature and emissivity evolution with redshift, loss of contrast against the CMB), thus making [CII] a reasonable alternative option to explore the gas content of QGs.

\begin{figure*}[tbh]
\centering
\includegraphics[scale=0.6]{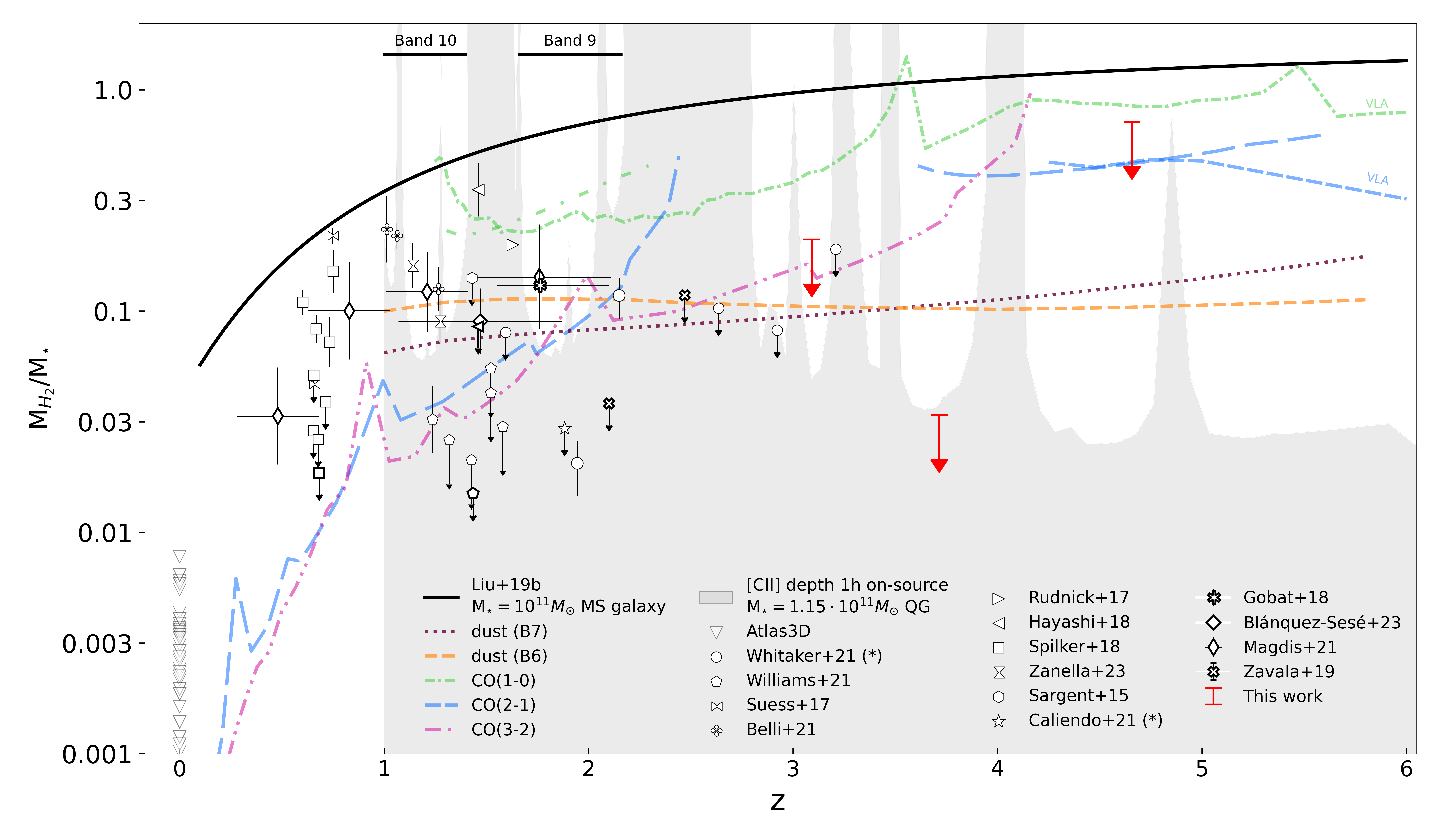}
\caption{Depths (3$\sigma$) of different gas mass tracers with redshift with 1 hr of on-source time (including ALMA Bands 1 and 2 or VLA, see Sect.~\ref{sec:Pros}). For clarity, the depth reached by the [CII] line is underlined by a grey shaded region. The black solid line shows the gas fraction evolution for a main sequence galaxy with M$_{\star}=10^{11}$ M$_{\odot}$ from \citealt{Liu19}. QGs from the literature are shown as white markers. Measurements from stacked samples are shown with thick black contours. Gas fraction values were converted to a Chabrier IMF wherever necessary. 
All CO measurements were homogenised to $\alpha_{\rm{CO}} $=4.4 M$_{\odot}$ (K km s$^{-1}$ pc$^{2}$)$^{-1}$. All dust measurements were homogenised to a G/D ratio of 92.
(*) Original measures from \citealt{Whitaker21b} have been corrected according to the stellar profile-based extraction in \citealt{Gobat22} and by using the dust template by \citealt{Magdis21}.}

\label{fig:codust}
\end{figure*}

Here we argue that the multi-phase origin and striking brightness of [CII] can be capitalised to probe the gaseous component of faint high-z QGs. This can be assessed in Fig.~\ref{fig:codust}, where we quantify and compare the depths on f$_{\rm{mol}}$ reached by [CII], CO(1-0)/(2-1)/(3-2) and dust continuum for a $M_{\star}\sim 1.2\times 10^{11} M_{\odot}$ \citep[$FWHM=670\,km s^{-1}$][]{Esdaile} galaxy at $z>3$ at fixed ALMA/VLA on-source time (1h) and fixed spatial resolution (0.4" beam). For CO transitions, we converted the rms of the ALMA Observing Tool (OT)\footnote{Or that of the VLA Exposure Calculator whenever necessary.} over 670 km s$^{-1}$ at different $\nu_{\rm{obs}}$ into L$^{\prime}_{\rm{CO}}$ using eq. (2) in \citealt{silverman18}. We then converted to M$_{\rm{mol}}$ by adopting $\alpha_{\rm{CO}} $=4.4 M$_{\odot}$(K km s$^{-1}$ pc$^{2}$)$^{-1}$ and average excitation ratios from nearby post-starburst galaxies \citep{French23}, likely appropriate for high-z QGs \citep{Deuge20}. For dust continuum, we compared the continuum-rms of the ALMA OT, sampled at the central $\nu_{\rm{obs}}$ of bands 6 and 7, with the corresponding flux density of the QGs dust template from \citealt{Gobat18} rescaled at a given z and assuming a gas-to-dust ratio (G/D) of 92~\citep{Magdis21}. Being a relative comparison, here we neglect the effect of the CMB. As can be seen, [CII] yields by far the best performance across almost the entire range from z=3 to 6 for non-lensed galaxies, reaching deeper limits than what allowed by dust continuum (2-4$\times$ deeper) and CO(2-1)/(1-0) (13-30$\times$ deeper). This translates into faster observations at fixed targeted f$_{\rm{mol}}$: up to 16$\times$ faster than dust continuum and 170-1000$\times$ faster than CO(2-1) and CO(1-0), respectively. We note that by using excitation ratios of regularly star-forming galaxies at z$\sim$1.6 \citep{Daddi15} the depths of CO(2-1) and CO(3-2) would improve by 50\%, but would still not go below f$_{\rm{mol}}\sim10$\% at z$\gtrsim$3 (CO(3-2)). As a result, these numbers would reduce the need for strongly lensed sources to probe gas fractions down to few per cent in the high-z universe. Interestingly, [CII] would perform equally well as CO(3-2), CO(2-1) and dust continuum (band 7) at z$\sim$1.8, which would serve as a sweet spot to calibrate the amount of L$_{\rm{[C II]}}$ arising from cold gas on lower redshift QGs.

Lastly, should [CII] observations reveal robust detections, it would open to dedicated follow-ups aimed at accessing the kinematics of cold gas in QGs in order to compare it to that of their stellar component from JWST IFU spectroscopy in search of potential signatures of past merger events or external accretion, if any \citep{Kalita22, Raimundo23}. Additionally the gas kinematics could be compared to that of plausible progenitors at higher redshifts \citep{Rizzo20, Lelli21}.

\section{Archival ALMA data}
\label{sec:selection}
The ALMA archive contains three high-z QGs with publicly available coordinates and [CII] coverage: ADF22-QG1 (Programme: 2013.1.00159.S , PI: H. Umehata), ZF-COS-20115 (Programme: 2015.A.00026.S , PI: C. Schreiber) and GS-9209 (Programme: 2015.1.01074.S , PI: H. Inami).

ADF22-QG1 is a massive QG at $z_{spec}=3.092^{+0.008}_{-0.004}$ \citep{Kubo21}, located at the core of the SSA22 protocluster \citep{Umehata15}. Its stellar mass, size, and Sersic index have been constrained to M$_{\star}= 1.66-3.72 \cdot 10^{11}$ M$_{\odot}$,  r$_{e}=1.01\pm$0.04 kpc (0.13"), and n$_{Sersic}=2.5\pm0.2$,  respectively \citep{Kubo17, Kubo21}. No measurement on its stellar velocity dispersion is currently available. The integration on this source is 4.5 min, thus likely insufficient for our purposes.

ZF-COS-20115 is located at $z_{spec}=3.715^{+0.002}_{-0.002}$ and its stellar mass, size, Sersic index, and stellar velocity dispersion are M$_{\star}=1.15^{+0.16}_{-0.09}\cdot 10^{11}$ M$_{\odot}$,  r$_{e}=0.66\pm$0.08 kpc (0.09"), n$_{Sersic}=4.1\pm$1.0, and $\sigma_{\star}=283\pm$52 km s$^{-1}$, respectively \citep{Schreiber18a,Esdaile}. ZF-COS-20115 has been extensively studied as it is part of a pair with a moderately star-forming companion galaxy (Hyde), undetected in H band, located at a distance 3.1 kpc \citep[inferred from the 0.43" offset from the HST and ALMA images and the $\Delta z=0.008$;][hereafter S18a and S18b respectively]{Schreiber18b, Schreiber18a}. Hyde is detected in [CII] emission showing a very broad, double-peaked profile associated to a rotating disk (v$_{rot}=781^{+218}_{-366}$ km s$^{-1}$, S18a) and offset from the QG by -550 km s$^{-1}$ in frequency space. Its [CII] spectrum is shown in blue in Fig.~\ref{fig:spectra}. The integration on this source is 1.2 hours.

GS-9209 is located at $z_{spec}=4.6582^{+0.0002}_{-0.0002}$ and its stellar mass, size, Sersic index and stellar velocity dispersion are M$_{\star}=4.1^{+0.2}_{-0.2}\cdot 10^{10}$ M$_{\odot}$,  r$_{e}=215\pm$20 pc (0.033"), n$_{Sersic}=2.3\pm$0.3, $\sigma_{\star}=247\pm$16 km s$^{-1}$, respectively \citep{Carnall23}. The integration on this source is 52 s.

The QGs considered here have been spectroscopically confirmed with an accuracy of $\delta z/(1+z)\sim 0.001$, 0.0004, and 0.00004, respectively. This translates into an uncertainty on the position of the [CII] line of $\pm$440 km s$^{-1}$, $\pm$159 km s$^{-1}$, and $\pm$11 km s$^{-1}$, respectively. None of the galaxies is resolved at the resolution of these datasets (beam sizes of 0.27"$\times$0.26", 0.54"$\times$0.43", and 0.19"$\times$0.16", respectively).

The QGs considered here have rest-frame optical spectral properties consistent with being post-starbursts. This category can contain misclassified dusty starbursts, galaxies still transitioning towards full quiescence and recently quenched but rather passive systems. The most robust indicator of any dust-obscured star formation activity is their IR-based SFRs \citep[SFR$_{IR}$,][]{PW00,Baron22}.
ADF22-QG1 has an SFR$_{IR}<$9-20 M$_{\odot}/yr$ \citep[$\sim$19 times below the coeval MS,][]{Schreiber15}, based on its 1.2mm flux \citep{Kubo21}. ZF-COS-20115 has an SFR$_{IR}<$13 $M_{\odot} /yr$ ($\sim$40 times below the coeval MS) based on its ALMA Band 8 continuum flux, conservatively assuming a dust temperature range of 20-35K \citep{Schreiber18b}. No SFR$_{IR}$ is currently available for GS-9209. The most stringent constraint available comes from the narrow component of its H$\alpha$ emission line \citep{Carnall23}, which yields SFR$_{H_{\alpha}} = 1.9 \pm 0.1 M_{\odot} /yr$ ($\sim$50 times below the MS at its redshift). Although these SFRs are inherently model dependent (ADF22-QG1 and ZF-COS-20115) or biased towards short wavelengths (GS-9209), we note that appropriate ALMA coverage for high-z QGs is still missing.

\section{Analysis and results}
\label{sec:obs}

\begin{figure}[tbh]
\centering
\includegraphics[scale=0.35]{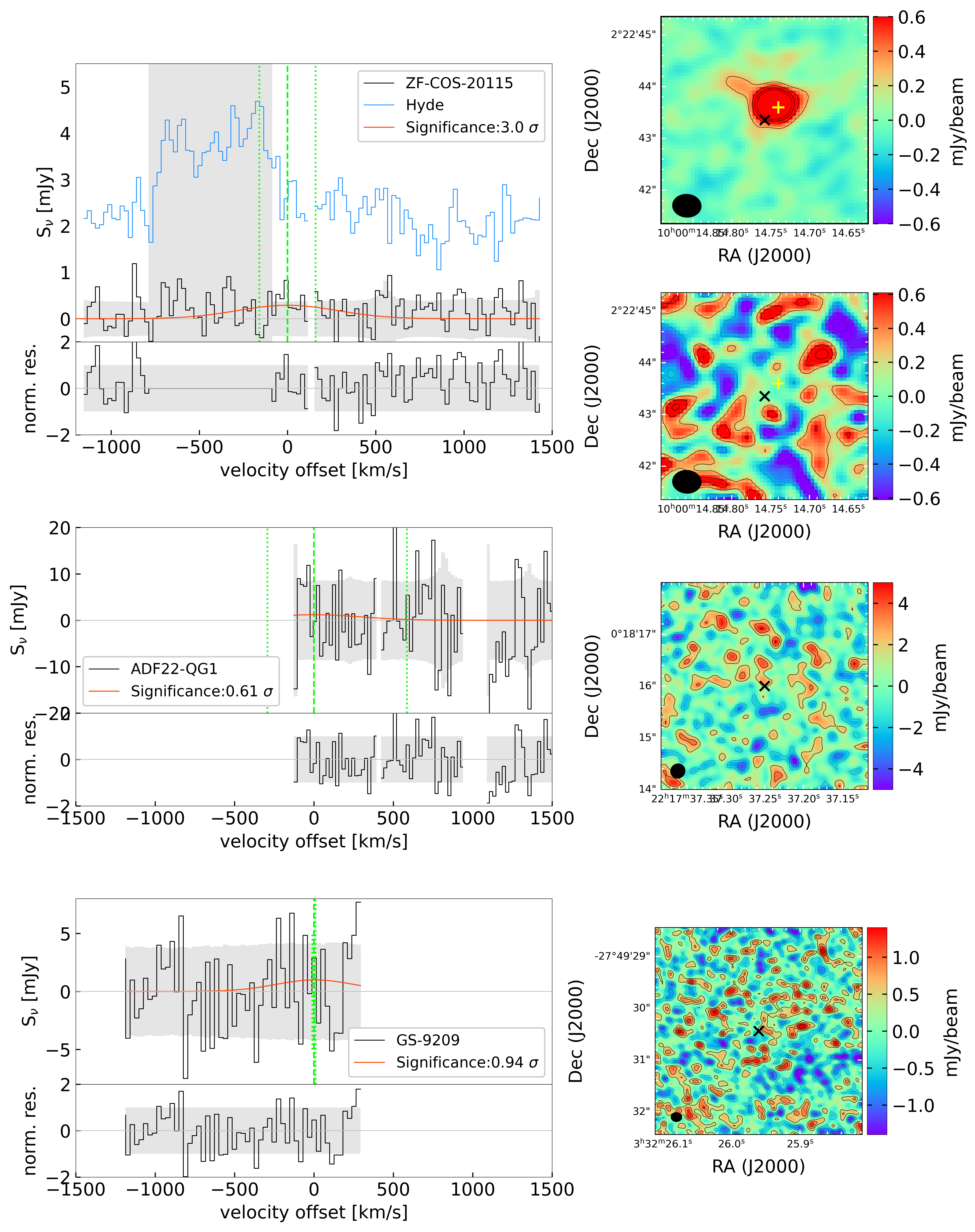}
\caption{[CII] coverage for our sources. Left column: Spectra extracted at the position of our targets. Dashed and dotted green lines mark, respectively, the expected $\nu_{obs, [CII]}$ and its uncertainty. Red curves mark MPFIT Gaussian fits at fixed frequency and FWHM. Channels where a significant contribution from Hyde could be expected were masked, as shown by the grey vertical area.
Right column: Dirty images averaged over the stellar FWHM of our sources: ZF-COS-20115, before and after Hyde's subtraction (top two panels); ADF22-QG1 (mid panel); GS9209 (bottom panel). The positions of the QGs are marked by black crosses. Contours show 1, 2, 3, 4 times the RMS of the image.}
\label{fig:spectra}
\end{figure}

We reduced the ALMA data following the standard CASA calibration script. The calibrated visibilities were converted to UVFITS data to allow for uv-plane source fitting with the UV\_FIT task of the GILDAS/MAPPING software package\footnote{\url{https://www.iram.fr/IRAMFR/GILDAS/}}, used for the spectral extraction. 
Whenever necessary, we imposed the target's coordinates as the phase centre and we corrected the extracted spectrum for the corresponding primary beam (PB) attenuation, as described below.
No detection was present in the spectra of our three galaxies (Fig.~\ref{fig:spectra}). We computed upper limits from the velocity integrated RMS taken as the RMS measured from the whole spectrum, multiplied by the square root of the number of channels within the expected FWHM of the line and by the channel width in km s$^{-1}$. For ZF-COS-20115 and GS-9209 their FWHM was fixed to the value arising from their high-resolution stellar spectra i.e. 665 km s$^{-1}$ and 580 km s$^{-1}$, respectively. We kept a conservative FWHM of 705 km s$^{-1}$ for ADF22-QG1 based on the latest measurements of \citealt{Forrest2022} ($\sigma_{\star}\sim$ 300 km  s$^{-1}$) for similarly massive high-z QGs.
For each source, we first converted the integrated RMS into L$_{\rm{[C II]}}$ and then into M$_{\rm{mol}}$ by adopting the $\alpha_{\rm{[CII],mol}}=31 M_{\odot}/L_{\odot}$ from Z18. We computed upper limits on f$_{\rm{mol}}$ by conservatively adopting the lowest stellar mass value for each source given the uncertainties quoted in the respective reference papers. 


\paragraph*{\textbf{ADF22-QG1:}}
The spectrum of this galaxy was extracted using a Spergel profile with parameters fixed to the published properties of the source (see Sect.~\ref{sec:selection}): Spergel index $\nu$=0.0 (corresponding to a Sersic index of 2.5), size of $r_{e}=0.13"$ and position angle of -5 \textdegree \citep{Kubo17}. The centre of its ALMA pointing is offset by 5.6" in the north-west direction compared to the position of the target. At the expected observed-frame frequency of [CII] the half power beam width (HPBW) is 13.6" from which we obtain a PB correction of $\times$1.6 applied to the extracted spectrum.
As expected, the sensitivity reached for ADF22-QG1 only results in a 0.6 significance. The corresponding 1$\sigma$ velocity integrated RMS is S$_{\rm{[CII]}} $<1.10 Jy km s$^{-1}$, that translates into a 3$\sigma$ upper limit of M$_{\rm{mol}}$<5.2 $\cdot 10^{10} M_{\odot}$ and a molecular gas fraction of f$_{\rm{mol}}$<21\%.
Converting the value of the rms outside the expected line channels divided by the square root of the number of channels, we obtain 1.05 mJy. This value is about eight times higher with respect to the dust SED from \citealt{Gobat18} rescaled to a z$_{spec}=$3.092 and to the same stellar mass.

\paragraph*{\textbf{GS-9209:}}
Due to its very compact nature, its spectrum was extracted as a point source.  
The centre of its ALMA pointing is offset by 6.5" in the south-west direction compared to the position of the source, from which, given the HPBW of 17.3", we obtain a PB correction of $\times$1.5 applied to the extracted spectrum.
The 1$\sigma$ velocity integrated RMS over the expected frequency of [CII] is S$_{\rm{[CII]}} $<0.467 Jy km s$^{-1}$, which translates into a 3$\sigma$ upper limit of M$_{\rm{mol}}$< 2.8 $\cdot 10^{10} M_{\odot}$. This corresponds to a molecular gas fraction upper limit of f$_{\rm{mol}}$<72\%. 

\paragraph*{\textbf{ZF-COS-20115:}}
At the position of the QG, the ALMA spectrum is still dominated by the star-forming neighbour. We tested whether any signal could be ascribed to the QG by subtracting Hyde's [CII] emission (and underlying continuum) in the uv plane. We fitted the quiescent and star-forming components simultaneously, by fixing the coordinates of the QG and letting the position of the star-forming galaxy vary in each channel. 
We adopted several models for Hyde's subtraction: elliptical Gaussians, exponential profiles and elliptical Spergel profiles with varying parameters (source size and position angle). The spectrum of the QG was extracted using the parameters listed in Sect.~\ref{sec:selection} with a Spergel profile of index $\nu$=-0.6 corresponding to n$_{Sersic}$=4.
The detection remains only tentative (significance levels ranging from 0.7$\sigma$ to 3.0$\sigma$) since the noise properties of the residual spectrum are dependent on Hyde's subtraction scheme. The highest significance is reached using a Spergel profile for ZF-COS-20115 and an elliptical Spergel profile for Hyde. 
As our main interest is to provide a proof of concept, we compute our gas fraction upper limit from the spectrum with the largest RMS (S$_{\rm{[CII]}} $<0.0775 km s$^{-1}$) and using the corresponding integrated value to place a 3$\sigma$ upper limit. 
We obtain M$_{\rm{mol}}$< 3.6$\times10^{9}$ $M_{\odot}$ corresponding to f$_{\rm{mol}}$<3.2\% assuming that the entire [CII] luminosity arises from cold molecular gas (see Fig.~\ref{fig:codust}). 
The upper limit increases to f$_{\rm{mol}}$<4.5\% when adopting the largest value of $\alpha_{\rm{[CII],mol}}$ allowed by the uncertainties in Z18.
We note that this is one of the most stringent individual constraints on the molecular gas mass of QGs at z$>$2 to date. The ALMA band 8 continuum RMS outside of the expected [CII] line ($\sim$0.04 mJy/beam) yields a dust continuum-driven f$_{\rm{mol}}$ upper limit of $\sim$15\%, five times shallower than what can be done with [CII] at fixed time investment, further supporting the method proposed here.\\

We explicitly note here that we assumed that the [CII] emission shares the same spatial distribution and velocity dispersion as the stars. [CII] halos more extended than the stellar component, however, have been found in several samples of star-forming galaxies at high redshift. We note that the beam sizes for ZF-COS-20115 and GS-9209 already reach 2.5-3 times their rest-frame optical r$_{e}$, which is the typical ratio of the [CII] sizes to the optical/UV sizes of both ALPINE and REBELS galaxies \citep{Fujimoto20, Fudamoto22}. Extracting the spectrum of ZF-COS-20115 over the entire beam (using a point source model) leaves our results unchanged. Applying UV tapering (tapered beam FWHM$\approx0.93\times 0.86"$, i.e. $ \approx 2 \times 3 \times r_{e}$) and performing a point source extraction for ADF22-QG1 still leads to an undetection with consequent upper limit of 50\% due to the increase in noise.

Moreover, although molecular gas may not strictly follow the stellar kinematics, we note that large line widths are conservative and that these upper limits scale with velocity interval as $\sqrt{\Delta v}$ \citep{Williams21}.
We also note that by performing Gaussian fitting on each spectrum with MPFIT \citep{Markwardt09}\footnote{\url{https://www-astro.physics.ox.ac.uk/~cappellari/software/}} at fixed expected observed-frame frequency and FWHM, imposing the continuum level to zero and rescaling the error spectrum to obtain a reduced $\chi^2$ of unity, our upper limits increase by 20\%, leaving our conclusions unaffected.

Lastly, we note that the the spectral coverage at our disposal for ADF22-QG1 and GS-9209 is incomplete. This prevents us from ruling out the potential presence of outflows that would offset the [CII] emission line from the near UV/optically derived z$_{spec}$ \citep[e.g.][]{Carniani17, Fujimoto20}.
Ruling out any potential presence of dust-obscured star formation and/or the presence of strong outflows will require additional targeted ALMA observations.

\section{Discussion and conclusions}
\label{sec:discussion}

In this letter we proposed a new method for constraining the cold gas budget of high-z QGs by extending to this galaxy population the use of the [CII] emission line, so far routinely used for star-forming objects. We then explored the ALMA archive to do the exercise of applying the L$_{\rm{[C II]}}$-to-H$_2$ conversion by Z18 on those high-z QGs covered with already existing [CII] observations.
Since this is the first time that such a method is adopted for QGs, we here mention caveats that make the conversion not straightforward and we discuss the implications of a low strict upper limit for ZF-COS-20115.

\subsection{Caveats}
\label{sec:Caveats}

\paragraph*{\textbf{The $\alpha_{\rm{[CII],mol}}$ conversion factor:}}
In Fig. 9 in Z18, the $\alpha_{\rm{[CII],mol}}$ conversion factor appears to be fairly independent on the sSFR of galaxies down to $\sim$0.3 sSFR/sSFR$_{\rm{MS}}$. For less star-forming or quenched objects, however, the points systematically shift to higher values of $\alpha_{\rm{[CII],mol}}$. If the trend is real, $\alpha_{\rm{[CII],mol}}$ might increase by a factor of 2-3 for galaxies ten times below the MS, leading to a systematic underestimation of integration times or molecular gas fractions at fixed depth when adopting the average $\alpha_{\rm{[CII],mol}}$ from Z18.  
In order to test whether an $\alpha_{\rm{[CII],mol}}$ significantly higher than 31 M$_{\odot}$/L$_{\odot}$ would be more appropriate for QGs, a suitable experiment is to follow-up with deep [CII] observations QGs already detected in either dust continuum or CO transitions. This would allow to have a first constraint on which L$_{\rm{[C II]}}$ would correspond to the previously derived molecular gas mass (M$_{H_2}$), albeit being still dependent on $\alpha_{\rm{CO}} $ or on the G/D conversion factors. Non-detections would be an indication that $\alpha_{\rm{[CII],mol}}$ would need to be revised for QGs or that a low G/D ratio would apply \citep{Donevski23}. As can be seen in Fig.1, QGs with M$_{H_2}$ measurements all lie at z$<$2.5. Those at 1.0$<$z$<$1.4 and 1.68$<$z$<$2.17 would thus be ideal targets because they are in redshift windows for which [CII] coverage is still provided by ALMA bands 10 and 9, respectively.

\paragraph*{\textbf{Contribution to L$_{\rm{[C II]}}$ by ionised and atomic gas:}}
In the case of detections, whether ionised gas has a larger contribution to L$_{\rm{[C II]}}$ in high-z QGs will need to be assessed. One solution would be to target the [NII] emission line at 205 $\mu$m, which has the same critical density as [CII]$_{158}$ but higher ionisation potential \citep[11.3 vs 14.5 eV][]{CW13}. Comparing the observed [CII]$_{158}$/[NII]$_{205}$ line ratio with the expected theoretical ratio if both [CII] and [NII] originated from purely ionised regions, it is possible to obtain a direct measure of the fraction of [CII] arising from the neutral medium  \citep[atomic and molecular, L17;][]{Schreiber22}. This line ratio is relatively insensitive to the electron density (n$_{e}$) and can vary, at most, by a factor of 1.6 over 3 dex in n$_{e}$. We caution, however, that current theoretical values available in the literature are mostly dependent on the assumed n([CII])/n$_{e}$ and n([NII])/n$_{e}$ gas-phase abundance \citep[][L17]{Oberst06}. Competitive integration times to reject ionised fractions higher than 40-50\% can be reached with [NII] at z$>$2.9, when the [NII]$_{205}$ emission line enters in ALMA band 7, thus compensating for the lower luminosity of [NII] with a gain in sensitivity \citep[e.g.][]{Schreiber22}. The same experiment would require strongly lensed galaxies at 1$<$z$<$2 due to the drop in sensitivity of ALMA bands 8 and 9.\\
On the contribution to L$_{\rm{[C II]}}$ from atomic gas: pinning down the H$_{2}$/HI mass ratio at high redshift, as well as the precise relative contribution of atomic and molecular gas to L$_{\rm{[C II]}}$, is a non-trivial task as both fractions are still uncertain \citep{OR09, Vallini15, Vizgan22b,Chowdhury2022}. The molecular-to-atomic gas mass ratio likely depends on the spatial scales probed, with molecular gas expected to be dominant in the central regions of galaxies due to enhanced hydrostatic pressure, as observed in local star-forming galaxies \citep{BR06} and ETGs \citep{Serra2012, Babyk2023}. Due to the increasing stellar density of QGs with redshift, molecular gas may thus be expected to be the primary gaseous component when limiting observations over spatial scales comparable to stellar sizes. Further studies are needed to accurately decompose the mass contribution of atomic versus molecular gas (and their mass-to-light ratio with respect to L$_{\rm{[C II]}}$) over different spatial scales.
Finally, we recall here that the uncertainty on the H$_{2}$/HI mass ratio affects the G/D ratio as well, being formally a tracer of the total gas M$_{gas}$=M$_{HI}$+M$_{H_{2}}$.

\paragraph*{\textbf{Effects of the CMB:}}
The increasing CMB temperature is expected to heat the gas of galaxies and to introduce an additional continuum component that would reduce the contrast of both dust continuum and line emissions. \citealt{Vallini15} and \citealt{Lagache18} argue that the effect of the CMB on the cold neutral medium (CNM, n<50 cm$^{-3}$) should be negligible up to z=4.5. Beyond this redshift, as the CMB temperature would rapidly reach the excitation temperature of [CII] at z=6.5 in the CNM, the recoverable [CII] emission from this component would be $<$10\% and only [CII] from PDRs would be recovered. The limiting redshift depends on the exact excitation temperature of [CII] in different ISM components, and hence on the kinetic temperature and density of the gas and the source of ionisation.
Following \citealt{dacunha13} and \citealt{Vallini15}, we compute the fraction of recoverable [CII] flux against the CMB for our galaxies. We conservatively assume the excitation temperature of the CNM from \citealt{Vallini15} \citep[T$_{\rm{ex, CNM}}$=22K, not too distant from T$_{\rm{dust}}$=21K in][]{Magdis21}. The recovered fractions for our galaxies are 0.98, 0.95 and 0.82 for ADF22-QG1, ZF-COS-20115 and GS-9209, respectively. Therefore, the CMB would have a non-negligible effect only on GS-9209, for which the upper limit would increase by 18\%.
Thus, when assuming that [CII] emission only arises from the CNM, this tracer would efficiently work up to z=4.5, with larger depth than other standard methods can reach with relatively short integration times on non-lensed galaxies.

\subsection{What can be inferred on quenching from deep upper limits}
\label{sec:sowhat}
Under the assumption that the $\alpha_{\rm{[CII],mol}}$ in Z18 applies to high-z QGs, an f$_{\rm{mol}}<$3\% for ZF-COS-20115 is $\sim$ 3 times lower than the extrapolation of trends reported in \citealt{Gobat18} and \citealt{Magdis21} (although within the scatter of \citealt{blanquez23}, see Fig.1). Such a low value reinforces the existence of extremely gas-poor objects at high-z, (as highlighted by previous works, e.g. \citealt{Whitaker21a}) and possibly of a large scatter in the gas content of the early quenched population. With 1.2 hours of integration, it highlights the daunting possibility that future [CII] observations might also yield non-detections, requiring large statistics and observational efforts to assess such scatter. Moreover, despite the galaxy being recently quenched, its gas fraction is 3-5 times lower than what has been observed for lower redshift post-starburst galaxies with a similar age \citep{Belli21, Suess17, Zanella23}. This implies that the initial gas fraction of newly quenched galaxies at high-z may not be constant and that additional parameters such as stellar mass may be needed when modelling the effect of progenitor bias into the gas fraction evolution of QGs \citep{Gobat20}.\\
According to the star formation history inferred for ZF-COS-20115 in S18b, the galaxy has been evolving passively for t$_{\rm{quench}}$=0.5$^{+0.2}_{-0.2}$ Gyr. 
An f$_{\rm{mol}}<$3\% after 0.5 Gyr from the last substantial drop in star formation implies that the galaxy did not experience substantial gas accretion after quenching and that bulge stabilisation \citep{Martig09} is not the main maintenance mechanism for its quiescence.
ZF-COS-20115 is associated to a moderately star-forming galaxy with M$_{\star}\sim 0.34-1.28 \cdot 10^{11} M_{\odot}$ and SFR$_{\rm{tot}}\sim 50^{+24}_{-18}$ M$_{\odot}$/yr, about a factor of 4 below the MS at its redshift \citep{Schreiber22}. Summing their stellar masses we obtain a maximum M$_{\star,tot}\sim 2.6 \cdot 10^{11} M_{\odot}$ which, assuming that the two galaxies belong to the same dark matter halo, leads to a maximum halo mass of $M_{\rm{DM}}\sim10^{13} M_{\odot}$, according to the stellar-to-halo mass relation at 3.5<z<4.5 in \citealt{Shuntov22}. At this redshift and halo mass, cold accretion in a hot medium is expected to take place \citep{DB06, Daddi22}. An absence of detectable gas down to $\sim$3\% in ZF-COS-20115, and a suppressed star formation activity in its companion galaxy, suggest that a mechanism other than virial shock heating is operating to prevent further accretion. This suggests that efficient heating of the surrounding circumgalactic medium to temperatures high enough to have a cooling time in excess of t$_{\rm{quench}}$ is needed. 
A viable internal mechanism to starve galaxies is AGN kinetic feedback \citep{Croton06}, that is qualitatively consistent with the increasing strength of radio-mode AGN activity in QGs towards higher redshifts \citep{deugenio21, Magdis21, ito22}. Deep radio observations are needed to quantitatively test this scenario.

In summary, we argue that at z>3 and for fixed ALMA time commitment, [CII] can push much deeper than other molecular gas tracers (see Fig.~\ref{fig:codust}) providing stringent constraints on which quenching processes are operating, thus representing an obvious choice for studying the gaseous component of QGs at high redshift. 


\begin{acknowledgements}
We thank the anonymous referee for their constructive comments. CDE is grateful to F. Rizzo, C. Circosta and I. Delvecchio for helpful discussions. CDE acknowledges funding from the MCIN/AEI (Spain) and the “NextGenerationEU”/PRTR (European Union) through the Juan de la Cierva-Formación program (FJC2021-047307-I).
This paper makes use of the following ALMA data: ADS/JAO.ALMA\#2013.1.00159.S, ADS/JAO.ALMA\#2015.A.00026.S and ADS/JAO.ALMA\#2015.1.01074.S. ALMA is a partnership of ESO (representing its member states), NSF (USA) and NINS (Japan), together with NRC (Canada), MOST and ASIAA (Taiwan), and KASI (Republic of Korea), in cooperation with the Republic of Chile. The Joint ALMA Observatory is operated by ESO, AUI/NRAO and NAOJ.
\end{acknowledgements}

\nocite{Spilker18}
\nocite{Suess17}
\nocite{Rudnick17}
\nocite{Hayashi18}
\nocite{zavala19}
\nocite{Sargent15}
\nocite{Caliendo21}
\nocite{Belli21}
\nocite{Zanella23}

\bibliographystyle{aa}
\bibliography{references.bib}

\end{document}